\documentstyle{article}[12pt]

\begin{document}
\Large
\title{Fermion families, and chirality through extra
dimensions}  
\author{Recai Erdem$^0$}
\maketitle
\footnotetext{E-mail:erdem@likya.iyte.edu.tr}
\small
\begin{center}
Department of Physics\\
Izmir Institute of Technology\\
G.O.P. Bulvari No.16\\
Izmir, Turkey
\end{center}
\normalsize
\begin{abstract}
We give a simple model to explain the origin of fermion families,
and chirality through the use of a domain wall placed in a
five dimensional space-time. 
\end{abstract}

\section{Introduction}

The popularity of higher dimensional spaces 
has revived in the last years by the hope that the extra space
dimensions other than the the usual three dimensions may be
experimentally accessible in near future. In this
context the idea of the confinement of the usual particle
spectrum into a four dimensional topological defect of the higher
dimensional space-time is also revived especially in the view
that by using Randall-Sundrum [1] type spaces one can
confine gravity as well in infinite dimensions. 
In this
study we shall consider a metric with a 4-dimensional
Poincare invariance [2] and a domain wall in a Randall-Sundrum
type space. We find that it gives some important clues towards
the understanding of the origin of fermion families and
chirality. By using a 
Minkowski-like metric
we get a domain wall solution in five dimensions where fermions may be
concentrated at more than one point of the wall. Each
concentration point corresponds to a fermion
family. After adding a fermion mass term to this scheme the left-handed
and right-handed fermions become concentrated at different regions in
the wall. In other words the wall itself acts as a mother 3-brane
which carries two sub-branes, right-handed and left-handed. Each
of these two sub-branes contain n different
symmetrical sub-subbranes whose locations can be identified with
different masses of different fermion families. Finally we include gravity
into the picture by using a Randall-Sundrum-like metric.
The framework employed here has some conceptual similarities with
the study of Dvali and Shifman, which considers families as
neighbors in a multi-brane world in five dimensions [3]. Another
study with some similar aspects is by Arkani-Hamed and Schmaltz
[4] where fermion mass
hierarchies$^1$\addtocounter{footnote}{1}\footnotetext{ In fact
the idea of
a better undertanding of fermion masses and mixings through higher
dimensional spaces is not new. Among these, the studies in the context of
heterotic string orbifolds seem to be especially appealing [5].}
are explained by the exponentially suppressed overlap of
fermion wave
functions located at different points in the extra
dimension(s). We will give a comparison of these studies
with the present one at the end of the next section after we give
the essential points of this study.

\section{Basic Lines of The Model}

Consider the following well-known 5-dimensional scalar Lagrangian
[6]
\begin{equation}
L=\frac{1}{2}\partial_A\phi\partial^A\phi-\frac{1}{4}\lambda(\phi^2-
\frac{\mu^2}{\lambda})^2~,~~~A=0,1,2,3,4 \label{a1}
\end{equation}
with 
\begin{equation}
 ds^2=\eta_{AB}dx^Adx^B=dx_\mu\;dx^\mu-(3ay^2+b^2)^2dy^2~,~~\mu=0,1,2,3
\label{a2}
\end{equation}
where $a$, $b$ are some constants and $(\eta_{AB})=diag(1,-1,-1,-1,-1)$ 
The equation of motion corresponding to 
Eq.(\ref{a1}), and its domain and anti-domain wall
solutions [7] are
\begin{eqnarray}
&&(\frac{1}{3ay^2+b})\frac{\partial}{\partial y}[
(\frac{1}{3ay^2+b})\frac{\partial}{\partial 
y}\phi]+\mu^2\phi-\lambda\phi^3=0~, \label{a4} \\
&&\phi_{cl}=\pm \frac{\mu}{\sqrt{\lambda}}tanh\,
[\frac{\mu}{\sqrt{2}}(ay^3+by+c)]~;
\label{a5}
\end{eqnarray}
respectively.

We take the following fermion-scalar interaction
Lagragian$^2$\addtocounter{footnote}{1}
\footnotetext{A similar Lagrangian is considered in [8]. In fact 
one can identify $\phi$ in this equation as the gauge field
corresponding to a sixth dimension. This can be, for example, 
done by embedding this five dimensional space in a six
dimensional space studied by Manton [9] where instead of taking
the extra dimensions be compact one should only assume rotational
symmetry. In that case $B_5=\phi$ in Eq.(\ref{a6}) should be
replaced by $\Phi+\tilde{\Phi}$ of Ref.[9]. 
However for the sake of simpilcity we take this term arise from a
general scalar-fermion interaction term.}. 
\begin{eqnarray}
&&i\bar{\Psi}\Gamma^\mu\,D_\mu\Psi+
i\bar{\Psi}\Gamma^4\,
\frac{1}{(3ay^2+b)}\frac{\partial\Psi}{\partial y}
+g\bar{\Psi}\phi\Psi \nonumber \\
&& \mbox{where}
~~\Gamma^4=-i\gamma_5~~,~~~D_\mu=\partial_\mu+igB_\mu
\label{a6}
\end{eqnarray}
In the presence
of the domain wall, the Dirac equation is 
\begin{equation}
i\gamma^\mu\,D_\mu\Psi+
\frac{1}{3ay^2+b}\gamma_5\frac{\partial\Psi}{\partial y}
+g\phi_{cl}\Psi=0 \label{a7}
\end{equation}
Assume that
\begin{equation}
i\gamma^\mu\,D_\mu\Psi=m
[\eta(y)-f]
\gamma_5\Psi \label{a8}
\end{equation}
where $\eta(y)$=$\frac{\mu}{\sqrt{2}}|(ay^3+by+c)|$ (the absolute value
results from the orbifold symmetry introduced in the next section).
This equation can be obtained, for example, by introducing an auxillary
fermionic field, $\chi$ coupling to the other fields only through the
following Lagrangian
\begin{equation}
L=\frac{1}{2}\partial_A\sigma\partial^A\sigma
+i\bar{\chi}\not{D}\Psi+
i\bar{\Psi}\not{D}\chi
-\alpha\bar{\chi}\sigma\gamma_5\Psi
-\alpha\bar{\Psi}\sigma\gamma_5\chi
\end{equation}
where
$\alpha=\frac{m\sqrt{\lambda}}{\mu}$ 
and $\not{D}=\gamma^\mu\,D_\mu$ . (One
should take $\alpha<<1$ in order to not conflict with phenomenlogy.
This is plausible because $\frac{m\sqrt{\lambda}}{\mu}$ is
a large number in general as is evident from Eq.(\ref{a144}) in the next
section.) After using Eq.(\ref{a8}), Eq.(\ref{a7}) reduces to
\begin{equation}
\frac{\partial\Psi_L}{\partial y}
-\frac{\partial\Psi_R}{\partial y}
+(3ay^2+b)(-\alpha\sigma_{cl}+g\phi_{cl})\Psi_R 
+(3ay^2+b)(\alpha\sigma_{cl}+g\phi_{cl})\Psi_L=0
\label{a9}
\end{equation}
where
$\Psi_L=\frac{1}{2}(1-\gamma_5)\Psi$,
$\Psi_R=\frac{1}{2}(1+\gamma_5)\Psi$.
The solutions of Eq.(\ref{a9}) are
\begin{eqnarray}
&&\Psi_R=exp[\beta\,ln(e^\eta+e^{-\eta})
\psi_R \nonumber \\ 
&&=(e^\eta+e^{-\eta})^{\beta}
e^{-m(\frac{1}{2}\eta^2-f\eta+d)]}
\psi_R
\nonumber \\
&&\Psi_L=
exp[-\beta\,ln(e^\eta+e^{-\eta})
-m(\frac{1}{2}\eta^2-f\eta+d)]\psi_L \nonumber \\
&=&(e^\eta+e^{-\eta})^{-\beta}
e^{-m(\frac{1}{2}\eta^2-f\eta+d)]}
\psi_L \nonumber \\
\label{a10}
\end{eqnarray}
where $\beta=g\frac{\mu}{\sqrt{\lambda}}$ and $\psi$ is the solution of
$i\gamma^\mu\,D_\mu\psi=m(\eta(y)-f)
\gamma_5\psi$.
The solutions are
well behaving provided $m>0$. 
One of
$(e^\eta+e^{-\eta})^{(\pm\beta)}$ is a monotonically increasing function
of $\eta$ and
the other is a monotonically decreasing function of $\eta$ while 
$e^{-m(\frac{1}{2}\eta^2-f\eta+d)]}$ is a monotonically decreasing
function provided $f<0$, $d>0$. Therefore 
$\Psi_L$ is a monotonically decreasing function of
$\eta$ (provided $\beta>0$) with its only maximum  
at $\eta=z_L=0$ while $\Psi_R$ has only one maximum at 
$\eta=z_R>0$.
We notice that $\Psi_L$ and $\Psi_R$ are concentrated at
different locations of the wall 
(that is at the maxima of $\Psi_L$ and $\Psi_R$) and they have different
distributions.
Moreover there are different fermions corresponding to different roots 
of $\eta=z_{L(R)}$ with exactly the same
probablity distrubutions at different locations of the wall. 
One can assume that al the roots of $z_L$ ($z_R$)
are much closer to each other than the roots of $z_R$ ($z_L$)
and the roots of $z_L$ are much closer to zero than the roots of $z_R$.
Then
all the fermion families corresponding to the roots of
$z_L$, that is $\Psi_L$'s, 
will be almost at the center of the 
wall while $\Psi_R$'s will be somewhat farther than the others. In
other words the probability of $\Psi_R$ participating in the
interactions whose gauge bosons well localised in the wall will
be signaficantly reduced. The number of the
roots of $z_L$, $z_R$ are three and can considered as locations
of different fermion families. 
Another interesting
aspect of the above
equations is that they give mass-like terms for chiral fermions while, 
as long as we are aware, the previous solutions are
given for massless chiral fermions [8,10]. The importance of this
construction is that one can embed this model in a six or higher
dimensional model. Then these mass-like terms will contribute to mass
matrix which will give the masses for the physical fermions after
diagonalization. In the case of six dimensions such a scheme will result
in the usual fermions and their mirrors with the same masses. So
physically
relavant models need to assign the fermions and their mirrors to different
gauge groups. In the case of seven or higher dimensional models it is
possible to give the fermions and their mirrors different masses provided 
the entries of the mass matrix are taken as general complex numbers. To
be more precise let us 
consider the following seven dimensional Dirac equation
\begin{eqnarray}
&&\Gamma^A\,D_A
\Psi=0 \\
&&\mbox{where}~~~~\Gamma^A\,D_A =
\left(\begin{array}{cc}
V&iD_5+D_6\\
iD_6-D_6&-V\end{array}\right) \nonumber \\
&&V=
i\gamma_5\,D_4+\gamma^\mu\,D_\mu
~~,~~D_A=\partial_A+igB_A ~~,~~A=0,1,...,6~,~~\mu=0,1,2,3 \nonumber
\end{eqnarray}
If either the gauge bosons
corresponding to extra dimensions have vacuum
expectation values or the derivatives give mass terms due to
compactification of the extra dimensions or due to both 
this equation induces a mass matrix 
\begin{eqnarray}
M=
\left(\begin{array}{cc}
i\gamma_5m_4&im_5+m_6\\
im_5-m_6&
-i\gamma_5m_4\\
\end{array}\right)
\label{a66} 
\end{eqnarray}
This equation leads to two different fermions, usual fermions and
their mirrors, with different masses
provided you take $m_i$'s as arbitrary complex numbers. Both masses 
become the same if one reduces the dimension of the space-time to
six or let all $m_i$'s be real. Another interesting feature of
the model is that some of the fermions can be localized in the
domain wall as some others are localized in the anti-domain wall,
corresponding to the same scalar field because the mass term 
in the exponent dominates over $\phi_{cl}$ to insure the
convergence of the argument of the exponent while this not
possible for the massless solutions of Dirac equation because in
that case both solutions can not be physical simultaneously. For example
if one
refers to Eq.(\ref{a10}) one notices (depending on the sign of
$\beta$) one of $\Psi_R$, $\Psi_L$ diverges as $m$ goes to zero such that
singling out one of them. This, in our opinion, 
may provide clue for why the right handed component of neutrinos are very
small. This result also gives an argument in favor of the unnaturalness of
exactly massless neutrinos.

As we have mentioned in the introduction the framework introduced
here has some conceptual similarities with the study of Dvali and
Shifman [3]. They simply assume that
there are more than one brane in the fifth dimension
without giving an explicit model which realizes this. They take the extra
dimension to be compact
and they do not consider the effect of gravity. Because of
experimental constraints [11] these brains must be close in the
extra dimension if they all contain the standard model
particles. This makes neglecting the gravity difficult and the
stablization of these brains a more subtle question. Moreover they give
their analysis on general grounds. Of course this approach has some
advantages such as providing a general framework for future
studies. However we believe that the introduction of
a more specific scheme will be phenomenologically more
promising. 
As we have mentioned in the
introduction another study which takes different fermions differ
by their locations in the extra dimension(s) is given by
Arkani-Hamed and Schmaltz [4]. Although they do not study the
problem
of fermion familes and chirality their study has some common
conceptual points with the present one. However they also do not
consider the effect of gravity and their space is compact. 
Moreover the fermion masses for different fermions are
simlpy taken different to put them in different locations in the
extra dimension(s) while in the present model different fermion masses
naturally arise as a result of the non-trivial form of the domain
wall.
However the models by Ref.[3] and Ref.[4] are stronger than the present
model in one aspect; 
they obtain the fermion masses by using the technique of the
overlap of wave functions as we do not introduce a method 
to derive the fermion masses. One can employ the same 
technique to obtain the fermion masses in this model. We leave
this point open to facilate consideration of different 
options as well in future.

\section{Inclusion of Gravity}

After pointing out that $m\gamma_5$ contributes to fermion
masses in higher (e.g. 7 or higher) dimensions we return back
to five dimensions to consider the issue of including
gravity into the picture. 
In the previous section the issue of confinement of
gravity (at least at low energies) to the usual four dimensional
space-time is not considered. So the next step is
the modification of our metric given in Eq.(\ref{a2}) to a
Randall-Sundrum-like form
\begin{eqnarray}
 ds^2&=&e^{A}
dx_\mu\;dx^\mu -(3ay^2+b)^2e^B\,dy^2 \nonumber \\
&=&
e^{-2\eta(y)}
dx_\mu\;dx^\mu
-
(3ay^2+b)^2e^{-6\eta(y)-2tanh\eta(y)+\frac{2}{3}tanh^3\eta(y)}
dy^2
\label{a11}
\end{eqnarray}
where under the assumption of orbifold symmetry
\begin{equation}
\eta(y)=\frac{\mu}{\sqrt{2}}|(ay^3+by+c)|
\label{a12}
\end{equation}
Unlike the Randall-Sundrum model we take only one
brane, that is, the domain wall due to $\phi_{cl}$.
The
relavant action is 
\begin{equation}
S= \int
d^5x\sqrt{-G}(\Lambda+{\cal L}_{cl}) \label{a13}
\end{equation}
where ${\cal L}_{cl}$ stands for the Lagrangian of the
classical
fields $\phi_{cl}$ and $\sigma_{cl}$ given in Eq.(\ref{a11}), $G$
is the five dimensional metric tensor, and $R$ is the
five dimensional Ricci scalar, $\Lambda$ stands for the
cosmological constant in the bulk. We take $\phi_{cl}$, $\sigma_{cl}$ to
be the classical solutions of $\phi$ and $\sigma$  given in the
previous section. This is plausible if we
assume that $\mu$ in Eq.(\ref{a13}) is very small and the metric in
Eq.(\ref{a2}) is an approximation of Eq.(\ref{a11}) for a sufficiently
broad range of $y$ values where $\eta\simeq 0$. Another view may be to
consider the metrics in Eq.(\ref{a2}) and Eq.(\ref{a13}) be different the
particular forms of a time dependent metric at two different times. For
example one may multiply $A$ and $B$ in Eq.(\ref{a13}) by 
$\frac{1}{2}(1+tanh\,x^0)$. Then the metrics Eq.(\ref{a2}) and
Eq.(\ref{a13}) correspond to its value at 
$x^0\rightarrow -\infty$ and
$x^0\rightarrow +\infty$, respectively. Hence one can suppose that the
classical solutions $\phi_{cl}$ and $\sigma_{cl}$ are created at  
$x^0<<0$ and they survive at the present (
Eq.(\ref{a13})) where $x^0>>0$.
Of course it is preferable to get the
exact domain wall solution corresponding to the above metric to get a full
view of the model but the corresponding equations seem to be rather
diffucult to solve and one may need to inculde additional scalar fields to
the spectrum to satisfy the Einstein equations in this general case. This
makes the analysis even more complicated. In fact the analysis of the
model for a sufficiently wide range of $y$ values is enough to see the
essential points of the model.

After replacing $\phi_{cl}$ and $\sigma_{cl}$ in ${\cal L}_{cl}$ (which 
is given in Eq.(\ref{a11})) one notices
that 
${\cal L}_{cl}$
reduces to 
${\cal L}_{cl}$=$-
\frac{\mu^4}{2\lambda}[(1-tanh^2\eta)^2
+\frac{1}{2})]$.
The $55$ component of the energy-momentum tensor is
$\frac{\mu^4}{2\lambda}(\eta^\prime)^2[(1-tanh^2\eta)^2+1]
+(\eta^\prime)^2{\cal L}_{cl}
$=
$\frac{\mu^4}{4\lambda}(\eta^\prime)^2$ where $\eta^\prime$ denotes the
derivative of $\eta$ with respect to $y$. The
corresponding Einstein equations are
\begin{equation}
R_{MN}-\frac{1}{2}G_{MN}R=
\frac{1}{4M^3}
[G_{MN}(\Lambda+{\cal L}_{cl}) -
\partial_M\phi_{cl}\partial_N\phi_{cl}]
-\partial_M\sigma_{cl}\partial_N\sigma_{cl}]
\label{a14}
\end{equation}
Under the assumption of the metric given in (\ref{a11}) the
Einstein equations [12,13] are satisfied for all $y$ provided
\begin{equation}
\Lambda=0,~~~
,~~~~48\lambda M^3=\mu^2 \label{a144}
\end{equation}

Another point worth to mention is that
the particles are 
confined at the center of the domain wall due to gravity as
well. In order to see this more clearly let us study the equation of
motion for graviton zero modes. As in Ref.1 we write the metric tensor
with linearized quantum fluctuations included as
$g_{MN}=G_{MN}+h_{MN}$. $h$ can be written as
$h_{MN}=\epsilon_{MN}(y)\,e^{ip.x}$ where $p^2=m^2$. 
We know that $h_{\mu\nu}$ must have a massless mode
corresponding
to the usual gravity. So this graviton zero mode must be confined to the
brane in order to prevent any conflict with inverse square law of gravity.
For this purpose one must write the linearized equation of motion for
$h_{\mu\nu}$. We work in the gauge $\partial^\mu h_{\mu\nu}=h^\mu_\mu=0$
as in Ref1. We expand the four dimensional metric tensor as 
$g_{\mu\nu}$=$e^{-2\eta(y)}\eta_{\mu\nu}+h_{\mu\nu}$
and 
$g_{55}$=$-
(3ay^2+b)^2
e^{-6\eta(y)-tanh\eta(y)+\frac{1}{3}tanh^3\eta(y)}
+h_{55}$
where
$\eta_{\mu\nu}$ is the Minkowski metric tensor. The
equation of motion for $h_{\mu\nu}$ is 
\begin{eqnarray}
&&[\frac{m^2}{2}e^{2\eta(y)}
-\frac{1}{2(3ay^2+b)^2}
e^{6\eta(y)+tanh\eta(y)-\frac{1}{3}tanh^3\eta(y)}
\partial_y^2 \nonumber \\
&&
-\frac{\mu^4}{2\lambda}(\eta^\prime)^2[(1-tanh^2\eta)^2+1]]\epsilon_{\mu\nu}=0
\end{eqnarray}
This is equivalent to
\begin{equation}
[-\frac{1}{2}\partial_y^2+V(y)]
\epsilon_{\mu\nu}=0
\end{equation}
where 
\begin{eqnarray}
&&V(y)=\frac{1}{2}m^2e^{2\eta(y)}(3ay^2+b)^2
e^{-6\eta(y)-2tanh\eta(y)+\frac{2}{3}tanh^3\eta(y)} \nonumber \\
&&-(3ay^2+b)^2
e^{-6\eta(y)-2tanh\eta(y)+\frac{2}{3}tanh^3\eta(y)}
\frac{\mu^4}{2\lambda}(\eta^\prime)^2[(1-tanh^2\eta)^2+1] \nonumber \\ 
\end{eqnarray}
One notices that the
potential
$V$ is
attractive for for large $y$'s while it is repulsive for small
$y$'s. In this way one can account for why the gravitational
attraction is small in our universe while the graviton  zero-mode is
localized in the fifth dimension. Another property of the above potential
is that the massive modes are less localized and their localization peak
is farther to the center of the domain wall than that of the zero mode.

The Dirac
equation is
\begin{eqnarray}
&&ie^{\eta(y)}\gamma^\mu\,D_\mu\Psi
+\frac{1}{(3ay^2+b)}e^{3\eta(y)+tanh\eta(y)-\frac{1}{3}tanh^3\eta(y)}
\gamma_5\frac{\partial\Psi}{\partial y}
+g\phi_{cl}\Psi=0  \nonumber \\
\label{a121}
\end{eqnarray}
We assume that
\begin{equation}
ie^{\eta(y)}
\gamma^\mu\,D_\mu\Psi=m
[\eta(y)-f]
\gamma_5\Psi 
\label{a131}
\end{equation}
The $y$ dependence of $\Psi$ for this metric changes because 
the equation (\ref{a9}) changes into
\begin{eqnarray}
&&\frac{\partial\Psi_R}{\partial y}
-\frac{\partial\Psi_L}{\partial y}  
+(\eta(y)-f)(3ay^2+b)e^{-3\eta(y)-tanh\eta(y) 
+\frac{1}{3}tanh^3\eta(y)}
[\alpha\sigma_{cL} 
+g\phi_{cl}]\Psi_R \nonumber \\
&&+(\eta(y)-f)(3ay^2+b)e^{-3\eta(y)-tanh\eta(y)+\frac{1}{3}tanh^3\eta(y)}
[-\alpha\sigma_{cl} 
+g\phi_{cl}]\Psi_L=0 \label{a91}
\end{eqnarray}
The $y$ dependence corresponding to this equation is essentially 
the same, in its form, as in the previous section. We do not give
the explicit $y$ dependence of $\Psi$ here because it is too
long.  
The new mass term for fermions becomes
\begin{equation}
[\eta(y_i)-f]
e^{\eta(y_i)}
\,
\,m\,\gamma_5
~~,~~~i=1,2,3 \label{a141}
\end{equation}
where $y_i$ denotes the location of the $i$'th family in the fifth
dimension.
The masses of different families (after embeding the model in
a higher dimensional space) can be made differ
significantly in this case due to the additional
exponent. Notice this was not the case in the previuos section
because we have to take $y_i$'s close to each other due to
experimental constraints on the scale of the observable part 
of extra dimensions. The result that the fermion masses are larger for
larger values of $y$ may seem surprising at first sight in a Kaluza-Klein
picture. On the other hand it is evident from equation (\ref{a10}) that
the more massive fermions are much more localized in the fifth
dimension. So there is no conflict with Kaluza-Klein view of fermion
masses. In fact the result that very massive fermions are located much
farther than the usual fermions may explain the reason behind the small
mixture of massive fermions with low mass fermions.

\section{Conclusion}

We have seen that there is a considerable hope for explaining
the origin of fermion families and chirality by using domain
walls in extra dimensions. 
We think that one of the most important
virtues of the present model is that it reaches almost all of
its conclusions through explicit formula instead of a vague
picture.
However there is still a long way
to go to put this scheme in a more detailed phenomenological
model which can give realistic description of chirality and
fermion families in the context of standard model. Probably in
such a description one should take the gauge bosons
corresponding to weak interactions to be localized on the
sub-brane containing the left-handed brane while the gauge
bosons of the non-chiral interactions can freely propagate
over the whole (mother) 3-brane. Such
an attempt may need to embed this simple scheme in higher 
dimensions. Such an extension may be done by giving similar
constructions and solving the corresponding equations for
vortices [14] or other topological defects. Another, maybe
simpler, route to go is to take the topological defect in higher
dimension to be a domain-wall junction or similar
intersections of multi branes in higher
dimensions [15]. All
these points will be clarified by further studies in future.

I would like to thank Professor R.N. Mohapatra, Professor O. Pashaev, 
and Dr. D.A. Demir for reading the manuscript and for their valuable
comments. I would also like to thank Dr. M. Schmaltz for his valuable
comments in the early stages of this work.


\begin{thebibliography}{99}
\bibitem{} L. Randall and R. Sundrum,
{\it{Phys.Rev.Lett.}} {\bf{ 84}}\rm (1999) 4690 
\bibitem{} V.A. Rubakov, M.E. Shaposhnikov, {\it Phys.Lett.} 
{\bf 125B} (1983)139 
\bibitem{} G. Dvali and M. Shifman, 
{\it Phys.Lett.} {\bf B 475} 
(2000) 295 
\bibitem{} N. Arkani-Hamed and M. Schmaltz, {\it Phys.Rev.} {\bf
D 61} (2000) 033005 
\bibitem{} J.A. Casas and C. Munoz, 
{\it Nucl.Phys.} {\bf B 332} (1990) 189; \\
J.A. Casas, F. Gomez, and C. Munoz, 
{\it Phys.Lett.} {\bf B 292} (1992) 42 
\bibitem{} 
V.A. Rubakov, M.E. Shaposhnikov, {\it Phys.Lett.} {\bf 125B} 
(1983)136 
\bibitem{} R. Rajaraman, \it{ Solitons and Instantons} \rm
(North-Holland, Amsterdam, 1987)
\bibitem{} D.B. Kaplan, 
{\it Phys.Lett.} {\bf B 288} (1992) 342;\\
D.B. Kaplan and M. Schmaltz, {\it Phys.Lett.} 
{\bf B 368} (1996) 44 
\bibitem{} N.S. Manton, 
{\it{Nucl.Phys.}} {\bf{ B 158}}\rm (1979) 141 
\bibitem{} C.G. Callan, Jr., J.A. Harvey, 
{\it{Nucl.Phys.}} {\bf{ B 250}}\rm (1985) 427 
\bibitem{} C.D. Carone,
{\it{Phys.Rev.}} {\bf{ D 61}}\rm (2000) 015008 
\bibitem{} L. Randall and R. Sundrum,
{\it{Phys.Rev.Lett.}} {\bf{ 83}}\rm (1999) 3370; \\
T. Shiromizu, K. Maeda and M. Sasaki, 
{\it{Phys.Rev.}} {\bf{ D 62}}\rm (2000) 024012 
\bibitem{} J.E. Kim and B. Kyae, 
{\it Phys.Lett.} {\bf B 486} (2000) 165
\bibitem{} H.B. Nielsen and , P. Olesen, 
{\it{Nucl.Phys.}} {\bf{ B 61}}\rm (1973) 45 
\bibitem{} T. Nihei, \it{ Gravity localization with a domain wall
junction in six dimensions}, \rm Preprint hep-th/0005014; \\
N. Arkani-Hamed, S. Dimopoulos, G. Dvali and N. Kaloper, 
{\it{Phys.Rev.Lett.}} {\bf{84}}\rm (2000) 586 


\end{thebibliography}
\end{document}